\documentclass[pra,twocolumn,showkeys,superscriptaddress,floatfix]{revtex4-1}

\usepackage[T1]{fontenc}
\usepackage[latin9]{inputenc}
\usepackage{epsfig}
\usepackage{latexsym}
\usepackage{graphicx}
\usepackage{dcolumn}
\usepackage{amssymb}
\usepackage{amsmath}
\usepackage{bm}
\usepackage{mathrsfs}
\usepackage{bigints}
\usepackage{upgreek}
\usepackage{color}
\usepackage{longtable}
\usepackage{xspace}
\usepackage[squaren]{SIunits}
\usepackage{epstopdf}
\usepackage[normalem]{ulem}

\usepackage{verbatim} %for block comments

\definecolor{mygreen}{RGB}{20,148,20}

\usepackage{filemod}

\newcommand{\eq}[1]{Eq.~\eqref{eq:#1}}

\begin{document}

% \title{Atomic electric quadrupole transition on localized surface plasmon} %Old title
\title{Coupling of atomic quadrupole transitions with resonant surface plasmons}

\author{Eng Aik Chan}
% \affiliation{School of Physical and Mathematical Sciences, Nanyang Technological University, 637371 Singapore.}
\affiliation{Centre for Disruptive Photonic Technologies, TPI \& SPMS Nanyang Technological University, 637371 Singapore}%
\author{Syed Abdullah Aljunid}%
\affiliation{Centre for Disruptive Photonic Technologies, TPI \& SPMS Nanyang Technological University, 637371 Singapore}%
\author{Giorgio Adamo}%
\affiliation{Centre for Disruptive Photonic Technologies, TPI \& SPMS Nanyang Technological University, 637371 Singapore}%
\author{Nikolay I. Zheludev}%
\affiliation{Centre for Disruptive Photonic Technologies, TPI \& SPMS Nanyang Technological University, 637371 Singapore}%
\affiliation{Optoelectronics Research Centre \& Centre for Photonic Metamaterials, University of Southampton, Southampton SO17 1BJ, United Kingdom}

\author{Martial Ducloy}%
\affiliation{Centre for Disruptive Photonic Technologies, TPI \& SPMS Nanyang Technological University, 637371 Singapore}%
\affiliation{Laboratoire de Physique des Lasers, UMR 7538 du CNRS, Universit\'{e} Paris13-Sorbonne-Paris-Cit\'{e}      F-93430 Villetaneuse, France}
% \affiliation{School of Physical and Mathematical Sciences, Nanyang Technological University, 637371 Singapore.}

\author{David Wilkowski}
\email{david.wilkowski@ntu.edu.sg}
\affiliation{Centre for Disruptive Photonic Technologies, TPI \& SPMS Nanyang Technological University, 637371 Singapore}%
\affiliation{MajuLab, CNRS-UCA-SU-NUS-NTU International Joint Research Unit, Singapore.}
% \affiliation{Centre for Disruptive Photonic Technologies, TPI, Nanyang Technological University, 637371 Singapore}%
\affiliation{Centre for Quantum Technologies, National University of Singapore, 117543 Singapore}
%\date{\filemodprint{\jobname}~~File: \jobname}

%\date{Version 2, October 14, 2016}

\begin{abstract}
We report on the coupling of an electric quadrupole transition in an atomic vapor with plasmonic excitation in a nanostructured metallic metamaterial. The quadrupole transition at 685 nm in the gas of cesium atoms is optically pumped, while the induced ground state population depletion is probed with light tuned on the strong electric dipole transition at 852 nm. We use selective reflection to resolve the Doppler-free hyperfine structure of cesium atoms. We observe  a strong modification of the reflection spectra at the presence of metamaterial and discuss the role of the spatial variation of the surface plasmon polariton on the quadrupole coupling.
\end{abstract}



\maketitle

\section{Introduction}
% ====================
Surface plasmons polaritons (SPP) are collective excitations of light and electron at the metal interface \cite{raether1988surface}. Outside the metal, the electromagnetic field decays exponentially from the surface.  At resonance, the confinement of the SPP is at maximum and solely determined by the ohmic losses in metal. An increase of confinement goes hand-in-hand with an increase of the local density of states. Hence an atom, located at the vicinity of the surface, would enhance its interaction with the electromagnetic field. In this context, cooperative enhancement of the atomic transition have been reported with cold atoms \cite{stehle2011plasmonically}. In this previous work, fluorescence emission has been analyzed in polarization to subtract the bulk emission from the plasmonic driven emission. If the plasmonic mode could also be longitudinally confined with a substantial decrease of its volume, the light-matter interaction could be further enhanced. Ultimately, the strong coupling regime might be achieved, where coherent coupling of the transition is stronger than incoherent relaxation and reversible energy exchange can be operated. The strong coupling regime has been seen in a number of physical systems \cite{dintinger2005strong,chikkaraddy2016single}, but not yet in atomic vapor interacting with plasmonic nanostructures, despite its enormous interest, in particular for quantum technologies applications. One major challenge resides in maintaining the position of the atom within the mode volume at close vicinity of the metallic materials \cite{LukinPRBsingleemitter}. As a promising road toward mode volume reduction, the uniform metallic surface can be nanostructured. In this context, Fano-like coupling \cite{aljunid2016atomic} and tailoring of the atom-surface Casimir-Polder interaction \cite{chan2018tailoring} have been reported.

On experiments done so far, the plasmon-atom interaction was investigated through resonant coupling on an atomic electric dipole transition. Here, atoms behave like point-like scatterers. In contrast, a finite size effect of atoms is revealed using higher order electric multipole transitions. Interestingly, multipole transitions have different selection rules. Thus some of them are energetically isolated, and can be addressed individually, far-off-resonance from strong electric dipole transitions. However, since $a_0k_0\ll 1$, electric dipole forbidden transitions are very weak in vacuum. Here the Bohr radius $a_0$ is taken as the typical size of the atom and $k_0$ is the wavenumber in vacuum. At a material interface, the electromagnetic field is hybridized with surface modes, and its effective wavelength diminishes. Thus, electric dipole forbidden transitions can be enhanced due to higher electric field gradients, as was proposed for ideally conducting nanowires \cite{PhysRevA.62.043818}, and later, with plasmon modes at the vicinity of metallic nano-rods \cite{PhysRevA.85.022501,Shibata17} or in presence of a periodic nano-slits array \cite{deguchi2009simulation}.  Previous experiments reported enhancement of the cesium $6^2\textrm{S}_{1/2}-5^2\textrm{D}_{5/2}$ electric quadrupole transition in evanescent field using total internal reflection of light at the dielectric surface \cite{PhysRevLett.92.053001,TojoPhysRevA.71.012508}. More recently, Rivera and co-authors suggested two-dimensional materials with large confinement factors \cite{rivera2016shrinking}. In this extreme regime, all transitions could have similar oscillator strength, conducting to a deep modification of the excitation and emission spectrum of atoms which could be exploited for tests of the quantum electrodynamic theory in some regimes never obtained. %Furthermore, large confinement factors might favor multi-photonic emission instead of usual single photon emission. In this case, the emitted photons are entangled in energy like it occurs for any down frequency conversion processes. Thus atoms, or more generally any quantum emitters, embedded in confined plasmon, could be utilized as correlated photons emitters at high emission rates for application in quantum information.

This paper reports on the observation of electric quadrupole excitation of a cesium atomic vapor at the vicinity of a metallic metamaterial (MM) surface. The MM-vapor system is probed using selective reflection (SR) spectroscopy techniques \cite{MartialJPhys1991} in a pump-probe configuration \cite{chan2016doppler}. The experimental data are found to be in good agreement with a model developed in the mean-field approximation, where the MM is substituted by a homogenous material having the same far-field optical properties. Going beyond the mean-field approximation, we found that the coupling of SPP with atoms is important but spread over a large Doppler broadened profile. In Sec. \ref{Sec_Setup}, we discuss the experimental setup. In particular, a pump laser, tuned at $685\,$nm excites both a plasmonic resonance of the MM and the cesium $6^2\textrm{S}_{1/2}-5^2\textrm{D}_{5/2}$ electric quadrupole transition. The induced modification of the atomic ground state population is probed with a laser on the $6^2\textrm{S}_{1/2}\,-\,6^2\textrm{P}_{3/2}$ electric dipole transition at $852\,$nm. Both laser beams are normal to the surface leading to a Doppler-free signal. In Sec. \ref{Sec_Result}, we present a quantitative comparison of two SR signals, one extracted from a dielectric-vapor interface and the second coming from the dielectric-MM-vapor interfaces. The comparison indicated that the signal from MM does not show noticeable enhancement, in quantitative agreement with a model developed in the mean-field approximation. In Sec. \ref{Sec_Theoretical_analysis}, a model, beyond the mean-field approximation, shows that the expected plasmonic-driven excitation enhancement of the atomic oscillator strength is distributed over a larger Doppler broadened signal which leads to a moderate global offset of the SR signal. Conclusion and perspectives are given in Sec. \ref{Sec_conclusion}.

\section{Experimental Setup} \label{Sec_Setup}

\begin{figure}[t!]
   \centering
   \includegraphics[width = \linewidth]{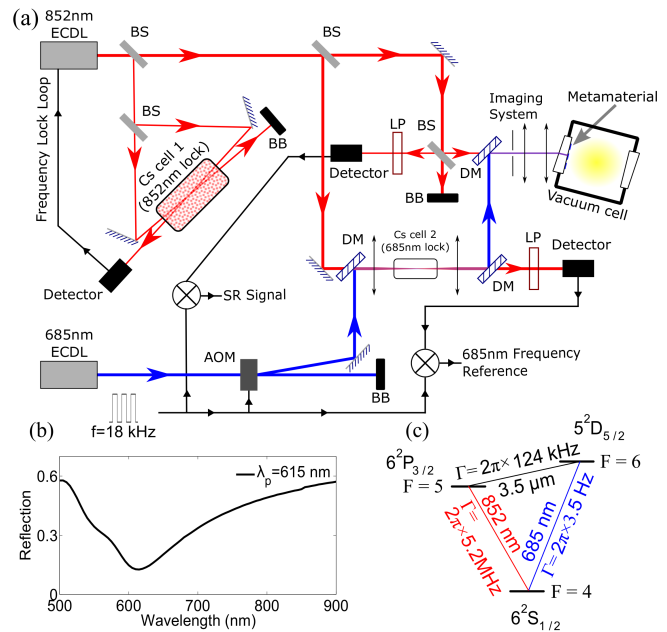}
   \caption{\label{Fig_1} (a) Sketch of the experimental setup (b) Reflection spectrum of the MM. (c) Cesium energy levels and transitions of interest. ECDL, external cavity diode laser; BS, beamsplitter; BB, beamblock; LP, longpass filter; DM, dichroic mirror; AOM, acousto-optic modulator}
\end{figure}

A general sketch of the experimental setup is shown in Fig. \ref{Fig_1}a. A vacuum chamber is brought to a temperature of $90\,^{\circ}$C, and filled up with cesium atoms at saturated vapor pressure. The typical atomic density is $N=8 \times10^{12}\,\textnormal{\,cm}^{-3}$. On a fused-silica viewport, a $50\,$nm thick layer of silver is deposed on which a periodic pattern of nano-slits is engraved (total area: $200 \mu m\times 200 \mu m$) The period and the unit cell structure are designed such that the MM host a fundamental resonant plasmonic mode at $615\,$nm, almost coinciding with the cesium $6^2\textrm{S}_{1/2}-5^2\textrm{D}_{5/2}$ electric quadrupole transition (see Fig. \ref{Fig_1}(b and c).

All the relevant spectra are obtained using a pump-probe technique, for more details see \cite{chan2016doppler}. In brief, a weak $852\,$nm probe laser is frequency locked on the $6^2\textrm{S}_{1/2},\,F=4-6^2\textrm{P}_{3/2},\,F=5$ hyperfine transition (cesium cell 1 in Fig. \ref{Fig_1}(a)). In addition, an amplitude-modulated $685\,$nm pump laser is scanned across the $6^2\textrm{S}_{1/2},\,F=4-5^2\textrm{D}_{5/2},\,F=6$ hyperfine transition, see Fig. \ref{Fig_1}(c). When the pump laser is at resonance, both lasers addressed the zero atomic velocity class. As a result, the ground state population is periodically decreased by the pump laser which modifies the probe laser interaction with the vapor. Thus, the pump amplitude modulation is transferred to the probe, providing a sensitive and ideally background-free demodulated signal of the ground state population. A first pump-probe setup is employed in transmission through a vapor cell for frequency reference purpose (cesium cell 2 in Fig. \ref{Fig_1}(a)). A second pump-probe setup performs SR spectroscopy on the atoms-MM system (vacuum cell in Fig. \ref{Fig_1}a). The two laser beams are co-propagating. The reflected beams go across a colored and a low-pass interference filter to remove the $685\,$nm beams and solely detect the $852\,$nm beam. In practice, a small fraction of the $685\,$nm light ($\sim 10^{-7}$ of the incident power) will also be detected and demodulated, which lead to an unwanted global offset of the SR signal.

Each laser beam has a waist of $150\,\mu$m which leads to a transit time of the atoms of around $1\,\mu$s (thermal velocity $\bar{v}\simeq 150\,\textrm{ms}^{-1}$). Since the transit time is about two times shorter that the $5^2\textrm{D}_{5/2}$ state natural lifetime, no Zeeman or hyperfine optical pumping, or Raman coherence effect need to be considered here.

 Using SR spectroscopy, we analyze the reflection properties of the probe laser at $852\,$nm at an interface between the cesium vapor and, either a dielectric, or the surface MM. The pump-probe setup allows us to observe the weak modification of the reflectance induced by excitation at $685\,$nm on the $852\,$nm beam.  A direct observation of the SR signal on the quadrupole transition, using a cesium vapor is possible but would require a higher atomic density \cite{PhysRevLett.92.053001,Aik_685_SR}. The pump-probe SR signal improvement is roughly given by the bare excited linewidth ratio $\Gamma(6^2\textrm{P}_{3/2})/\Gamma(5^2\textnormal{D}_{5/2})\simeq 40$. This signal improvement goes with a broadening of the SR signal imposed by the large dipole transition.

The observed relative modification of the reflectance is of the order of $10^{-8}$ and required one week of continuous integration to extract the weak signal out of the noise. Each scan of the $685\,$nm laser is performed in 200 ms and repeated at least $3\times 10^6$ times. To avoid frequency drifts of the $685\,$nm laser during acquisition, we record the pump-probe transmission signal of the bulk reference cell. The centre-of-mass of the spectrum is calculated in real time (at a rate of $5\,$s$^{-1}$) to find the mean laser frequency. A laser frequency drift is transposed into an error signal which is feedback to the piezoelectric ceramic controlling the laser cavity length.

\section{Selective reflection and mean-field interpretation} \label{Sec_Result}

\begin{figure}[t!]
   \centering
   \includegraphics[width = \linewidth]{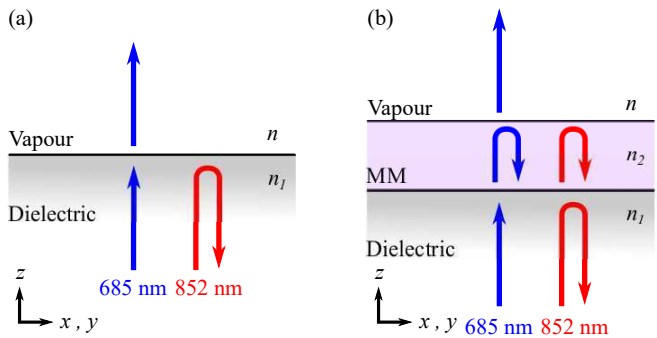}
   \caption{\label{Fig_3} Geometry of the laser beams at; (a) the dielectric-vapor interface, and (b) the dielectric-MM-vapor interface. In the mean-field approximation, the MM is replaced by a homogenous material which has the same index of refraction and attenuation coefficient. U-shaped arrows on the MM layer indicate that only a single reflection from the MM-vapor interface is considered because of large absorption in the MM (see text for more details).}
\end{figure}

\begin{figure}[t!]
   \centering
   \includegraphics[width =\linewidth]{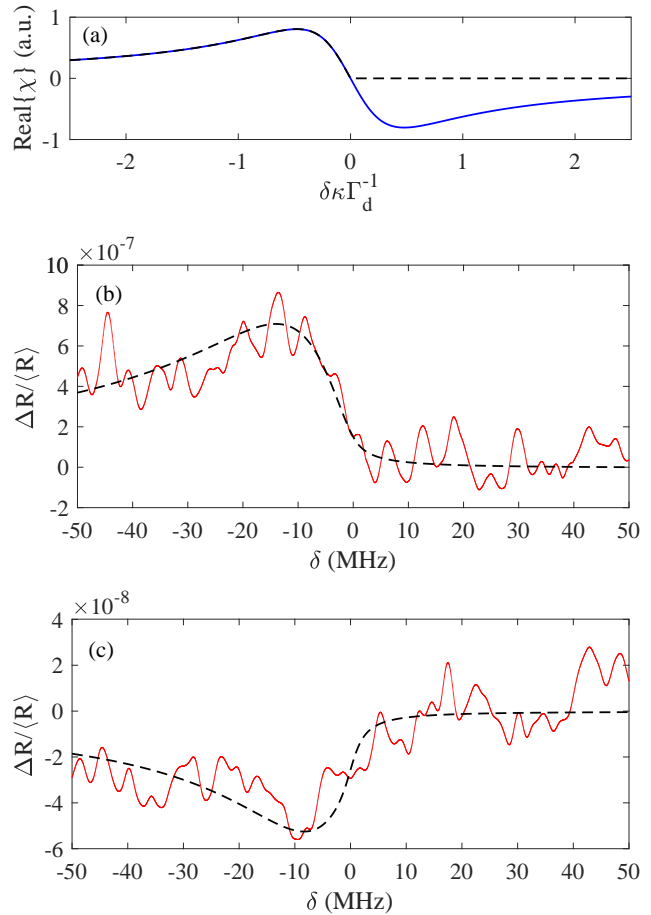}
      \caption{\label{Fig_2} Experimental (red) and theoretical (dashed black) curves as a function of the $685\,$nm laser frequency detuning. The frequency origin corresponds to the $6^2\textrm{S}_{1/2},\,F=4-5^2\textrm{D}_{5/2},\,F=6$ hyperfine transition. (a) Real part of the effective susceptibility given by Eq. (\ref{eq_chi}). The linewidth corresponds to $\Gamma_d/\kappa=6.5\,$MHz. The blue curve corresponds to the susceptibility of a bulk atomic ensemble at rest. (b) Relative reflectance. The experiment is performed on a dielectric-vapor interface. (c) Relative reflectance. The experiment is performed on a surface MM-vapor interface. The theoretical curves are given by inserting Eq. (\ref{eq_chi}) into Eq. (\ref{eq_R_DV}), and retaining only the demodulated term proportional to $I$. The $852\,$nm linewidth is taken at $20\,$MHz whereas the $685\,$nm linewidth is taken at $1\,$MHz (see text for more details). The effective index of refraction of the metasurface is $0.8$ and the attenuation coefficient is $4.05$. A same fitting parameter, for the overall amplitude of the theoretical curves, is used for both the dielectric, and MM cases. The SR signal is weaker for the MM, in agreement with the mean-field approximation. A lowpass filtering of the experimental data has been performed with a cut-off frequency of 1.2 MHz in frequency scan unit.}
\end{figure}

We first consider a single interface between a homogenous dielectric material and the atomic vapor, see Fig. \ref{Fig_3}a. To calculate the SR signal, one has to distinguish between atoms desorbing from the surface (velocity $v_z>0$) and atoms arriving from the bulk vapor ($v_z\leq0$). Ground-state atoms leaving the surface do not have time to undergo optical excitations mediated by the 685 nm laser. On the other hand, atoms arriving toward the surface, are in a steady-state regime. After some algebra, discussed in detail in \cite{SchullerDucloyPhysRevA.43.443,NeinhuisSchullerPhysRevA.50.1586}, one finds that the effective susceptibility of the vapor for SR of the probe at normal incidence reads :
\begin{align}
\chi_d(\delta)=i\frac{N\mu_{d}^2}{\epsilon_0\hbar\sqrt{2\pi}\bar{v}}\int_{-\infty}^{+\infty}\textrm{d}v_z
\frac{\Pi_g(v_z)\exp{\left(-v_z^2/2\bar{v}^2\right)}}{\Gamma_{d}/2-i(\delta_{d}+k_{d}|v_z|)},
\label{eq_chi_gene}\end{align}
where $\Pi_g(v_z)$ is the ground state ($6^2\textrm{S}_{1/2}$) population with,
\[ \Pi_g(v_z) =
  \begin{cases}
    1       & \quad \text{if } v_z>0\\
    1-\frac{I}{2I_s}\frac{\Gamma^2}{4(\delta-kv_z)^2+\Gamma^2}  & \quad \text{if } v_z\leq0.
  \end{cases}
\]
$N$ is the atomic density. $\epsilon_0$, $\hbar$, and $\bar{v}$ are the vacuum permittivity, the Planck constant, and the thermal velocity, respectively. $\mu$, $\Gamma$, $\delta$, and $k$ are the atomic dipole moment, the bare linewidth of the excited state, the frequency detuning and the laser wavenumber, respectively. If the subscript letter \emph{d} is attached to these quantities, we refer to the electric dipole transition $6^2\textrm{S}_{1/2}-6^2\textrm{P}_{3/2}$ (the probe transition). On the other hand, the absence of subscripts indicates that we consider the electric quadrupole transition $6^2\textrm{S}_{1/2}-5^2\textrm{D}_{5/2}$ (the pump transition). To simplify our theoretical approach, we disregard the Zeeman and hyperfine structure of the transitions. Moreover, we use the weak field limit where the laser intensity is below the saturation intensity. $I=TI_0\simeq 0.2\,$W/cm$^{2}$ is the modulated pump laser intensity and $I_s=2\,$W/cm$^{2}$ is the saturation intensity \cite{chan2016doppler}. $I_0$ is the incident laser intensity and $T=4n_1/(n_1+1)^2\simeq0.96$ is the transmittance at $685\,$nm of the dielectric-vapor interface. $n_1=1.45$ is the index of refraction of the dielectric. Since $\Gamma\ll\Gamma_d,k\bar{v},k_d\bar{v}$, we replace the factor $1/(4(\delta-kv_z)^2+\Gamma^2)$ by $\pi\delta_D(\delta-kv_z)/2\Gamma$, coming from $\Pi_g(v_z)$ in Eq. (\ref{eq_chi_gene}). Under this approximation the susceptibility takes the following analytical form,
\begin{equation}
\chi_d(\delta)=\chi_{d(0)}-iN\frac{\sqrt{2\pi}\mu_{d}^2}{8\epsilon_0\hbar}\frac{I}{I_s}\frac{\Gamma}{ k_d\bar{v}}\frac{1}{\Gamma_d/(2\kappa)+i\delta}\mathrm{\Theta}(-\delta).
\label{eq_chi}\end{equation}
$\delta_D(x)$ is the Dirac $\delta$ function, $\mathrm{\Theta}(x)$ is the step function, and $\kappa=k_d/k=0.8$. Here the probe laser at $852\,$nm is maintained at resonance, so $\delta_d=0$. $\chi_{d(0)}$ is the pump intensity independent part of the effective susceptibility which does not contribute to the demodulated SR signal. The effective susceptibility, depicted in Eq. (\ref{eq_chi}), has a simple physical interpretation. For $\delta\leq0$, it corresponds to the susceptibility of an atomic ensemble at rest coupled to the electric quadrupole transition with a bare linewidth of $\Gamma_d/\kappa$ due to the detection on the electric dipole transition. For $\delta>0$ the susceptibility is zero because of the relaxation of the population at the interface; see dashed black curve in Fig. \ref{Fig_2}(a). %As mentioned earlier, the asymmetry is due to the presence of the interface.

Since the medium is diluted, $|\chi_d|\ll1$, the vapor index of the refraction is $n\simeq 1+\chi_d/2$, and the reflectance of the probe is
\begin{equation}
R=|r_0|^2+2\textrm{Re}\{r_0^*\rho\chi_d\},
\label{eq_R_DV}
\end{equation}
where $r_0=(n_1-1)/(n_1+1)$ is the reflection coefficient without atomic vapor and $\rho=dr/d\chi_d|_{\chi_d=0}=-n_1/(n_1+1)^2$. The demodulated signal is proportional to the variation of reflectance, $\Delta R$, due to the atomic vapor and in the presence of the $685\,$nm pump laser. Since $r_0$, and $\rho$ are real quantities, the demodulated signal is simply proportional to the real part of $\chi_d$ (see Eq. (\ref{eq_R_DV})). In Fig. \ref{Fig_2}(b), we show the relative reflectance $\Delta R/\langle R\rangle$ of the probe laser at $852\,$nm as a function of the pump $685\,$nm laser frequency. $\langle R\rangle$ corresponds to the mean reflectance (DC component of the reflected signal). Since part of the reflected $685\,$nm laser is also detected, the demodulated signal shows a significant offset signal, which has been numerically removed. The experimental signal is in good agreement with the expected theoretical prediction (see black dashed curve on Fig. \ref{Fig_2}(b)). However, we have to artificially broaden the pump transition to $1\,$MHz, and probe transition to $20\,$MHz. These extra transition broadenings might be due to atom-atom collisions, transit-time broadening, and Casimir-Polder interactions, as was reported on similar studies \cite{chevrollier_high_1992,aljunid2016atomic,chan2018tailoring}, and might be due as well to finite pump laser linewidth.

In Fig. \ref{Fig_2}(c), we show the SR signal obtained for the dielectric-MM-vapor interface. Up to a sign factor, we observe similar features as for the dielectric-vapor interface, indicating that no further frequency broadening mechanisms are at play on the SR signal. We observe as well a substantial reduction of the signal amplitude of almost one order of magnitude.

To compare the experimental data with the model, developed above, for a dielectric interface, we carry out a mean-field approximation of the MM. It consists of replacing the MM by a homogenous material which has the same thickness and the same far-field transmittance and reflectance properties. Using finite-difference frequency-domain (FDFD) COMSOL simulations, we find that the homogenous material is characterized by a complex index of refraction of $n_2\simeq0.02+i3.65$ at the probe wavelength of $852\,$nm. The transmittance of the MM layer at $852\,$nm is predicted at $T_d=n_1\left|4n_2/\left[(n_1+n_2)(n_2+1)\right]\right|^2e^{(-2\textrm{Im}\{n_2\}k_dL)}=0.1$ in agreement with the measured value of $0.13$. Here, $L=50\,$nm is the MM thickness. Since $2\textrm{Im}\{n_2\}k_dL\simeq 2.7$, the absorption in the MM material is large and only one passage in the MM is considered (multi reflections can be ignored), see illustration in Fig. \ref{Fig_3}(b). We follow this approximation to calculate the reflection coefficient of the dielectric-MM-vapor interface and find $r_0=(n_1-n_2)/(n_1+n_2)$ and $\rho=-4n_1\left(n_2/\left[(n_2+1)(n_2+n_1)\right]\right)^2e^{(-2in_2k_dL)}$. At the pump frequency a plasmonic resonance is present leading to a large total transmission coefficient, measured to be $T\simeq0.77$. Following the same procedure as depicted earlier for the dielectric-vapor interface, we estimate the quantity $\Delta R/\langle R\rangle$, see black-dashed curves in Fig. \ref{Fig_2}(c). 

In the presence of MM, $r_0$ and $\rho$ are complex quantities, meaning that the SR signal, given by Eq. (\ref{eq_R_DV}), is a non-trivial mixture of the absorptive (\emph{i.e.} imaginary) and dispersive (\emph{i.e.} real) part of the atomic susceptibility. Moreover, in the mean-field approximation, the MM acts as an etalon. In contrast, for the dielectric-vapor interface, $r_0$ and $\rho$ are real quantities with no etalon effect. Those differences between the dielectric-vapor interface and the dielectric-MM-vapor interface explain the changes observed in the mean-field prediction for those two configurations; in particular, the opposite sign of the SR signals. More generally, using a dipole transition and MMs with different plasmonic resonances, it was shown that the coupling of the broad plasmonic resonance with the narrow atomic resonance lead to a Fano-like lineshape \cite{aljunid2016atomic,stern2014fano}. 

\begin{figure}[t!]
   \centering
   \includegraphics[width = \linewidth]{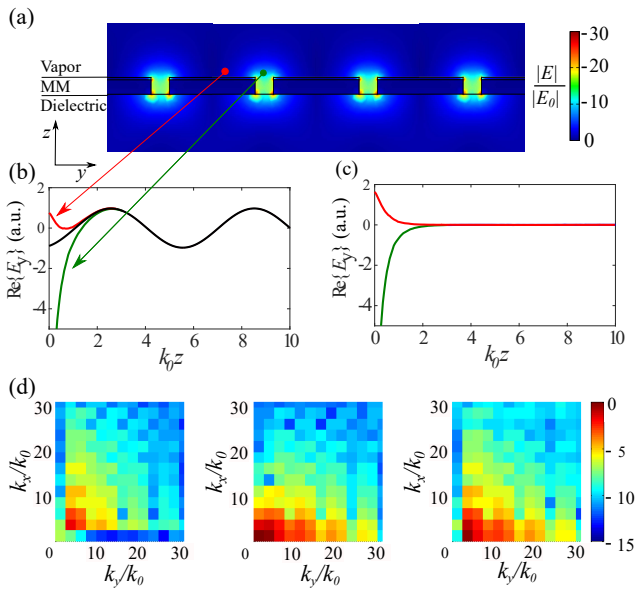}
   \caption{\label{Fig_4} (a) False-color image of the optical field intensity at the dielectric-MM-vapor interface along a plane normal to the MM cutting the MM at the center of the nano-slits. (b) \emph{y}-component of the electric field along the \emph{z}-axis for $z>0$. The incident polarization of the field is along \emph{y}. The green (red) curve corresponds to axis center (off-center) on the nano-slit. The black curve is the propagating contribution to the total electric field. (c) Same as (b), but the propagating contribution has been removed. Thus, the green and red curves show the SPP field. (d) Two-dimensional Fourier transform of the three components of the electric field, in logarithmic scale. The left (resp. middle, and right) panel represents the $\tilde{E}_x$ (resp. $\tilde{E}_y$, and $\tilde{E}_z$) Fourier component of the electric field. The SPP field is characterized by a wavenumber that is larger than $k_0$, the free space wavenumber. The propagation field corresponds to the origin of the $\tilde{E}_y$ component.}
\end{figure}

The mean-field approximation gives a good prediction of the SR signal on the dielectric-MM-vapor interface. At first sight, it seems a surprising result because the mean-field approximation washes out the SPP modes which give a large contribution to the electromagnetic field at the vicinity of the MM-vapor interface. Since the SR signal comes from atoms located within a short distance $k_d^{-1}$ from the interface \cite{aljunid2016atomic}, we would expect a large contribution of the SPP field to the SR signal as well. Moreover, the SPP modes are localized around the nano-slits which favor a strong electric field gradient for larger coupling to the quadrupole transition, see Fig. \ref{Fig_4}(a). However, we will show in Sec. \ref{Sec_Theoretical_analysis} that the SPP modes are characterized by a large wavevector that is parallel to the MM-vapor interface. Thus, their contribution to the atomic response is Doppler broadened which results in a moderated global offset of the SR signal, which is difficult to observe in the experiment. Therefore, only the propagating mode, which is normal to the interface, contributes to the SR signal. This propagating mode also corresponds to the far-field contribution of the MM, and it remains unchanged in the mean-field approximation. It explains why the mean-field approximation gives the correct prediction for our experiments.

\section{Surface plasmon polariton contribution} \label{Sec_Theoretical_analysis}
% ----------------------

As discussed in previous sections, we are investigating electric quadrupole coupling of atoms at the vicinity of a surface (within a distance $k_d^{-1}$) using a SR pump-probe experiment. The frequency of the pump is scanned across the $6^2\textrm{S}_{1/2},\,F=4-5^2\textrm{D}_{5/2},\,F=6$ electric quadrupole transition at $685\,$nm and modifies the atom ground state population at the resonant velocity class. The ground state population is probed on the $6^2\textrm{S}_{1/2},\,F=4-6^2\textrm{P}_{3/2},\,F=5$ electric dipole transition at $852\,$nm. For an incident beam at normal incidence, we found a Doppler-free SR signal, meaning that mainly atoms, flying parallel to the surface, contribute to the signal. We will now considering this situation with surface MM, ignoring the displacement of atoms normal to the surface (\emph{i.e.}, $v_z=0$).

Considering a MM located at the $z=0$, the electromagnetic field takes the following form, for $z>0$ (in the atomic vapor):
\begin{align}
    \vec{E}(x,y,z)&= E_0\cos{(k_0z-\omega t)}\hat{y}+\sum_{p,q\neq 0,0}\vec{E}^{(p,q)}(z)\nonumber\\
    &\times\cos{(pk_x^{SPP}x-\omega t+\phi_p)}\nonumber\\
    &\times\cos{(qk_y^{SPP}y-\omega t+\phi_q)}. \label{eq:spfield}
\end{align}
The electric field $\vec{E}(x,y,z)$ is computed thanks to a FDFD simulation. An example of the optical field intensity distribution is shown in Fig. \ref{Fig_4}a. We observe a strong enhancement of the field at the vicinity of the nano-slit. In Fig. \ref{Fig_4}(b), we show the electric field component Re$\{E_y\}$ at two different locations with respect to the nano-slit (red and green curves). Far from the surface ($k_0z>2$), the electric field reduces to an oscillating component (black curve). It corresponds to the propagating transmitted field and is described by the first right-hand side term in Eq.~(\ref{eq:spfield}). This field is polarized along the $y$-axis as the incident field. In Fig. \ref{Fig_4}(c), we remove the propagating components to reveal the near-field contribution of the SPP waves. They correspond to the second right-hand side term in Eq.~(\ref{eq:spfield}) where we perform a Fourier transformation in the $x-y$ plane parallel to the MM. The indices $p$ and $q$ label the Fourier components with amplitudes $\vec{E}^{(p,q)}(z)$, along the $x$ and $y$ axes, respectively. Because of the nano-slits, they are varying rapidly as shown in Fig. \ref{Fig_4}(d). Moreover, the SPP waves are stationary, and thus $\vec{E}^{(p,q)}(z)=\vec{E}^{(-p,-q)}(z)$. Finally, the components $\vec{E}^{(p,q)}(z)$ exponentially decay along the $z$-axis, normal to the MM, with a characteristic reciprocal length given by $k_{p,q}=k^{SPP}\sqrt{p^2+q^2}$. Here, $k^{SPP}=2\pi/\Lambda$ is the wavenumber of the fundamental modes of the SPP wave, where $\Lambda =284\,$nm is the length of the square unitary cell.

For an electric quadrupole transition, the square of the Rabi frequency takes the following form \cite{PhysRevA.85.022501}
\begin{equation}
\Omega^2=C^2\sum_{m}\left|\left\langle Y_2^m\left|\frac{\bar{\bar{Q}}}{r^2}\right|Y_0^0\right\rangle:\vec{\nabla}\vec{E}\right|^2,
\label{eq:Rabi_plamon}
\end{equation}
where $\bar{\bar{Q}}$ is the quadrupole moment operator, $\vec{\nabla}\vec{E}$ is the Jacobian matrix of the electric field, $r$ is the position operator, and $C$ is a constant coefficient taking into account the contribution to the coupling of the radial part of the electronic wave-function. The $Y_l^m$ are the spherical harmonics, where, to simplify the calculation, we have ignored the nuclear and electronic spin of the atomic state.

\begin{figure}[t!]
   \centering
   \includegraphics[width = \linewidth]{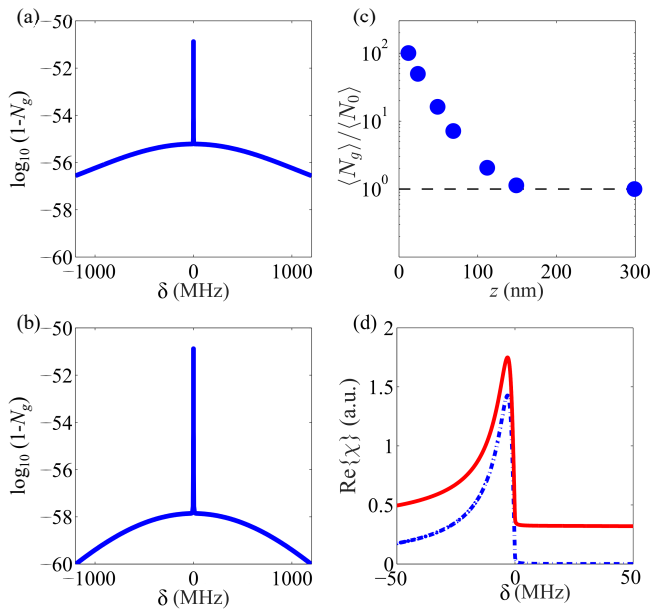}
   \caption{\label{Fig_5} (a) Spectrum of the ground state population depletion at a distance $z=13\,\textrm{nm}\simeq0.1k_d^{-1}$ from the MM-vapor interface. The temperature of the gas is $90\,^{\circ}$C and $\Omega_0=2 \times 10^{-11}\Gamma$. (b) Same as (a) but for  $z=70\,\textrm{nm}\simeq0.5k_d^{-1}$. The central narrow peak corresponds to the natural linewidth of 124 kHz. (c) Ground-state population depletion as a function of the distance from the MM-vapor interface. $\langle N_g\rangle$ is the spectral averaged ground state population. $\langle N_0\rangle$ is the spectral averaged ground-state population in vacuum. (d) Real part of the effective susceptibility. The red (resp. dashed-dotted blue) curve includes (resp. excludes) SPP.}
   % Omega_0 value derived from Codes\Comsol\Quad\Matlab Codes\grd_pop_fft5.m
\end{figure}

Computations of Eq.~(\ref{eq:Rabi_plamon}) have been reported for various nano-structured geometries \cite{PhysRevA.85.022501,Shibata17,deguchi2009simulation}. In those theoretical works, atoms are considered at rest and the Jacobian matrix $\vec{\nabla}\vec{E}$ is computed at the atom's position. In our case, we consider atoms flying above the MM which requires a time-dependent approach to this problem. However in the weak-field limit, all the contributions of the electromagnetic field Fourier components can be summed up independently. Thus we are able to come back to a simple time independent approach in the Fourier space. The finite velocity of the atoms simply leads to a frequency shift of the atomic resonance position. Doing so the mean ground state population, obtained after thermal averaging, reduces to
\begin{align}
    \bar{\Pi}_g &=1-\frac{\Omega_0^2}{4\delta^2+\Gamma^2}-\frac{1}{2\pi\bar{v}^2}\int\textrm{d}v_x\textrm{d}v_y\exp{\left(-\frac{v_x^2+v_y^2}{2\bar{v}^2}\right)}\nonumber\\
    &\times\sum_{p,q\neq 0,0}\frac{\Omega^2_{p,q}}{4(\delta-pk_x^{SPP}v_x-qk_y^{SPP}v_y)^2+\Gamma^2}. \label{eq:twolevel_plamon}
\end{align}
where $\Omega_0$ is the Rabi frequency from the propagating transmitted field, and $\Omega_{p,q}$ is the Rabi frequency from the SPP wave with Fourier components $p,q$.

Since the thermal distribution is a smooth function within the atomic resonance lorentzian profile, Eq. (\ref{eq:twolevel_plamon}) can be further simplified as such,
\begin{align}
    \bar{\Pi}_g &\simeq1-\frac{\Omega_0^2}{4\delta^2+\Gamma^2}-\sqrt{\frac{\pi}{2}}\frac{1}{2\Gamma\bar{v}}\sum_{p,q\neq 0,0}\frac{ \Omega^2_{p,q}}{k_{p,q}}\exp{\left(-\frac{\delta^2}{2k_{p,q}^2\bar{v}^2}\right)}.
    \label{eq:twolevel_plamon_approx}
\end{align}
The second right-hand side term comes from the propagating field. It gives the SR contribution as discussed in Sec. \ref{Sec_Result}. The last right-hand term comes from the SPP field not included in the mean-field approximation.

We note that the lifetime of the excited state is mainly dominated by the $5^2\textrm{D}_{5/2}-6^2\textrm{P}_{3/2}$ spontaneous emission decay time at $3.5\,\mu$m, see Fig. \ref{Fig_1}. At this wavelength, the MM does not have plasmonic resonance so the transition linewidth $\Gamma$ is expected to be the same as in free space.

In Fig. \ref{Fig_5}(a) and (b), we show the spectra of the ground-state population depletion at $z=13\,$nm and $z=70\,$nm from the MM-vapor interface using Eqs.~(\ref{eq:twolevel_plamon_approx}) and (\ref{eq:Rabi_plamon}). The Fourier components of the SPP field, extracted from FDFD simulation, are used to compute $\Omega_{p,q}$. The spectra are characterized by a narrow central peak with a \emph{z}-independent amplitude confirming that it can be attributed to the Doppler-free propagating component of the light field. The SPP field gives a Doppler broadened contribution of width $1.4 \times 10^4 \Gamma= 1.7 \,$GHz (FWHM value). The SPP field decays rapidly with the distance $z$ as illustrated in Fig. \ref{Fig_5}(c), where we show the frequency integrated ground state depletion as function of the distance $z$. Nevertheless we observe large enhancement of the coupling if the atoms are located close enough to the interface. In our experiment, the SR signal is roughly the averaged contribution of the atomic layers up to a distance of $k_d^{-1}=135\,$nm. To get a general idea of the SPP contribution to the SR signal, we plug the ground state depletion given by  \eq{twolevel_plamon_approx} into Eq. (\ref{eq_chi_gene}) for $z=70\,$nm. We observe that the general shape of the effective susceptibility remains unchanged, and the SPP field mainly contributes to an extra background, see Fig. \ref{Fig_5}(d). The moderate offset, expected on the SR signal, makes the observation of the SPP field very difficult to isolate from other spurious sources of offset as the stray reflection of the pump laser.

\section{Conclusion} \label{Sec_conclusion}

We have studied the spectral response of an electric quadrupole transition close to a material surface by using a pump-probe spectroscopy method. Here the SR signal is probed on a coupled electric dipole transition. On a dielectric-vapor interface, the SR signal is proportional to the susceptibility of an atom at rest for negative frequency and it drops to zero for positive frequency. The effective linewidth of the SR signal is predicted to be the linewidth of the probe transition divided by the probe and pump wavenumbers ratio. On a dielectric-MM-vapor interface, we observe a similar SR signal with a reduction of the overall signal amplitude. Here, the MM is designed to have a plasmonic resonance at the vicinity of the pump frequency. The overall signal reduction is in good agreement with a mean-field approximation where the MM is replaced by a homogenous material. The model indicates that the enhancement of the oscillator strength of the electric quadrupole transition, due to large spatial variations of the SPP field, is not observed on the SR signal. Using a complete description of the electromagnetic field, we find that the SPP waves contribute to a Doppler broadened signal which slightly offsets the SR signal. One way to remove the Doppler broadening would be to transversally cool the atomic vapor. In this case, a large increase of the SPP field coupling to the quadrupole transition should be observed.

This work is supported by the Singapore Ministry of Education Academic Research Fund Tier 3 (Grant No. MOE2011-T3-1-005 and MOE2016-T3-1-006(S)), and by the EPSRC UK Project No. EP/ M009122/1.

% Biblio
%\bibliographystyle{apsrev}
%\bibliographystyle{abbrv}
%\bibliography{biblio_plasmon_Q}
\newcommand{\noopsort}[1]{}\providecommand{\noopsort}[1]{}\providecommand{\singleletter}[1]{#1}%
\end{document}